\documentclass{iau} 
\usepackage{graphicx}
\usepackage{multirow}
\title[WD companions to BSSs in M67] 
{Detection of White Dwarf Companions to Blue Straggler Stars from UVIT Observations of M67}

\author[Sindhu et. al.]   
{Sindhu N$^{1,2}$,
Annapurni Subramaniam$^2$, Aaron M. Geller$^{3,4}$, Vikrant  Jadhav$^{2,5}$, Christian Knigge$^6$, Mirko Simunovic$^{7,8}$,   Nathan Leigh $^{9,10}$, Michael Shara$^{10}$ \and Thomas H. Puzia$^{11}$}

\affiliation{$^1$School of Advanced Sciences, VIT, Vellore -632014, India\\ email: {\tt sindhu.n@iiap.res.in} \\[\affilskip]
$^2$Indian Institute of Astrophysics, Koramangala II Block, Bangalore-560034, India\\[\affilskip]$^3$CIERA and Department of Physics \& Astronomy, Northwestern University, Evanston, IL 60201, USA\\[\affilskip]$^4$Adler Planetarium, Dept. of Astronomy, Chicago, IL 60605, USA\\[\affilskip]$^5$JAP and Physics Department, Indian Institute of Science, Bangalore - 560012, India\\[\affilskip]$^6$School of Physics and Astronomy, University of Southampton, Southampton, UK\\[\affilskip]$^7$Gemini Observatory, 670 N Aohoku Pl, Hilo, Hawaii 96720\\[\affilskip]$^8$Subaru Telescope, 650 N Aohoku Pl, Hilo, Hawaii 96720, USA\\[\affilskip]$^9$Departamento de Astronom\'ia, Facultad de Ciencias F\'isicas y Matem\'aticas, Universidad de Concepci\'on, Concepci\'on, Chile\\[\affilskip]$^{10}$Department of Astrophysics, American Museum of Natural History, NY 10024, USA\\[\affilskip]$^{11}$Institute of Astrophysics, Pontificia Universidad Cat\'olica de Chile, Santiago, Chile}

\pubyear{2019}
\volume{351}  
\jname{Star Clusters: From the Milky Way to the Early Universe}
\editors{A. Bragaglia, M.B. Davies, A. Sills \& E. Vesperini, eds.}
\begin{document}

\maketitle

\begin{abstract}
We investigate the old open cluster M67 using ultraviolet photometric data of Ultra-Violet Imaging Telescope in multi-filter far-UV bands. M67, well known for the presence of several blue straggler stars (BSS), has been put to detailed tests to understand their formation pathways. Currently, there are three accepted formation channels: mass transfer due to Roche-lobe overflow in binary systems, stellar mergers either due to dynamical collisions or through coalescence of close binaries. So far, there had not been any confirmed detection of a white dwarf (WD) companion to any of the BSSs in this cluster. Here, we present the detection of WD companions to 5 bright BSSs in M67. The multiwavelength spectral energy distributions covering 0.12 -11.5 $\mu$m range, were found to require binary spectral fits for 5 BSSs, consisting of a cool (BSS) and a hot companion. The parameters (Luminosity, Temperature, Radius and Mass) of the hot companions suggest them to be WDs with mass in the range 0.2 - 0.35 M$_{\odot}$ with T$_{eff}$ $\sim$ 11000 - 24000 K.
\end{abstract}
\firstsection 
\section{Introduction}
Stars evolve from the main-sequence (MS) to the red giant phase after they exhaust hydrogen burning in their core. Blue straggler stars (BSSs) lie above the MS turn-off (MSTO) in a star cluster, where stars of similar age have evolved, suggesting that BSSs have continued to stay on the MS, defying further evolution. These stars 
are believed to have gained mass by some process, resulting in a rejuvenation, which is not well understood. Three formation mechanisms are currently widely accepted: stellar mergers due to dynamical collisions (\cite[Hills \& Day 1976]{Hills1976}), coalescence of close binaries either in a binary, triple or higher order systems (\cite[Perets \& Fabrycky 2009]{Perets2009}, \cite[Naoz \& Fabrycky 2014]{Naoz2014}), mass transfer (MT) due to Roche-lobe overflow in binary systems (\cite[McCrea 1964]{McCrea1964}). The dominant BSS formation mechanisms operating in star clusters are likely to be dependent on evolution of binary stars (\cite[Knigge et~al. 2009]{Knigge2009}; \cite[Mathieu \& Geller 2009]{Mathieu2009Natur}; \cite[Leigh \& Sills 2011]{Leigh2011}).

So far, only a few studies in star clusters have documented observational evidence for the formation of BSSs through MT channel by detecting a hot companion to the BSSs (\cite[Knigge et~al. 2008]{Knigge2008}; \cite[Gosnell et~al. 2015]{gosnell2015}; \cite[Subramaniam et~al. 2016]{2016Subramaniam} and \cite[Sahu et al. 2019]{Sneha2019}). In open clusters, MT channel of BSS formation is preferred by models, which suggests the presence of a hot companion to the BSS. The suggested remnants of the post-MT systems are white dwarfs (WDs), post-asymptotic giant branch (AGB), early- AGB, AGB-Manqu\'e, or post-horizontal branch (HB) stars, which are expected to be in the temperature range of 12,000 - 30,000 K. Some of these systems are hot low luminosity objects with small radii. In optical images, their presence is inconspicuous, due to the overwhelming flux from the BSS. 

In the open cluster M67, \cite[Milone \& Latham (1992)]{milone1992f190} estimated the companion mass of BSS (WOCS 1007) to be 0.21 -- 0.19 M$_\odot$ from the spectroscopic orbital solutions. This star was suggested as a BSS+WD by \cite[Shetrone \& Sandquist (2000)]{Shetrone2000}, due to the estimated mass of the companion. There has been no confirmed detection of WD companions to any of the BSSs in M67, except for a couple of suggestions.

\section{Observations \& Method}
The \textit{Ultra-Violet Imaging Telescope} \textit{(UVIT)} on {\it ASTROSAT}, the first Indian space observatory, has been producing images of $\sim$1.5 arcsec resolution in both far-ultraviolet (FUV) and near-UV (NUV) bands. M67 observations were performed on 23 April 2017 by UVIT. We obtained the FUV data in three filters, viz. F148W, F154W and F169M filters, all observed on the same day. The standard IRAF routines were followed to obtain the magnitudes in all three filter images, which are also corrected for aperture and saturation (\cite[Tandon et~al. 2017]{TandonCalb2017}) 

We combine the UVIT magnitudes with other estimations in the UV, optical and near-infrared (IR) from space as well as ground observations to create multiwavelength spectral energy distribution (SED), spanning a wavelength range of 0.12 - 11.5 $\mu$m. The SEDs are analyzed with the help of theoretical spectra to detect and characterise the BSSs as well as hot components, if it is present. We used VOSA to fit the SEDs and the details of fitting along with the steps involved for composite fit in case of presence of a binary companion are given in \cite[Sindhu et~al. (2018)]{Sindhu2018} and \cite[Sindhu et al. (2019)]{2019Sindhu}

\section{Analysis}
\begin{figure}
\begin{center}
\includegraphics[width=2.35in]{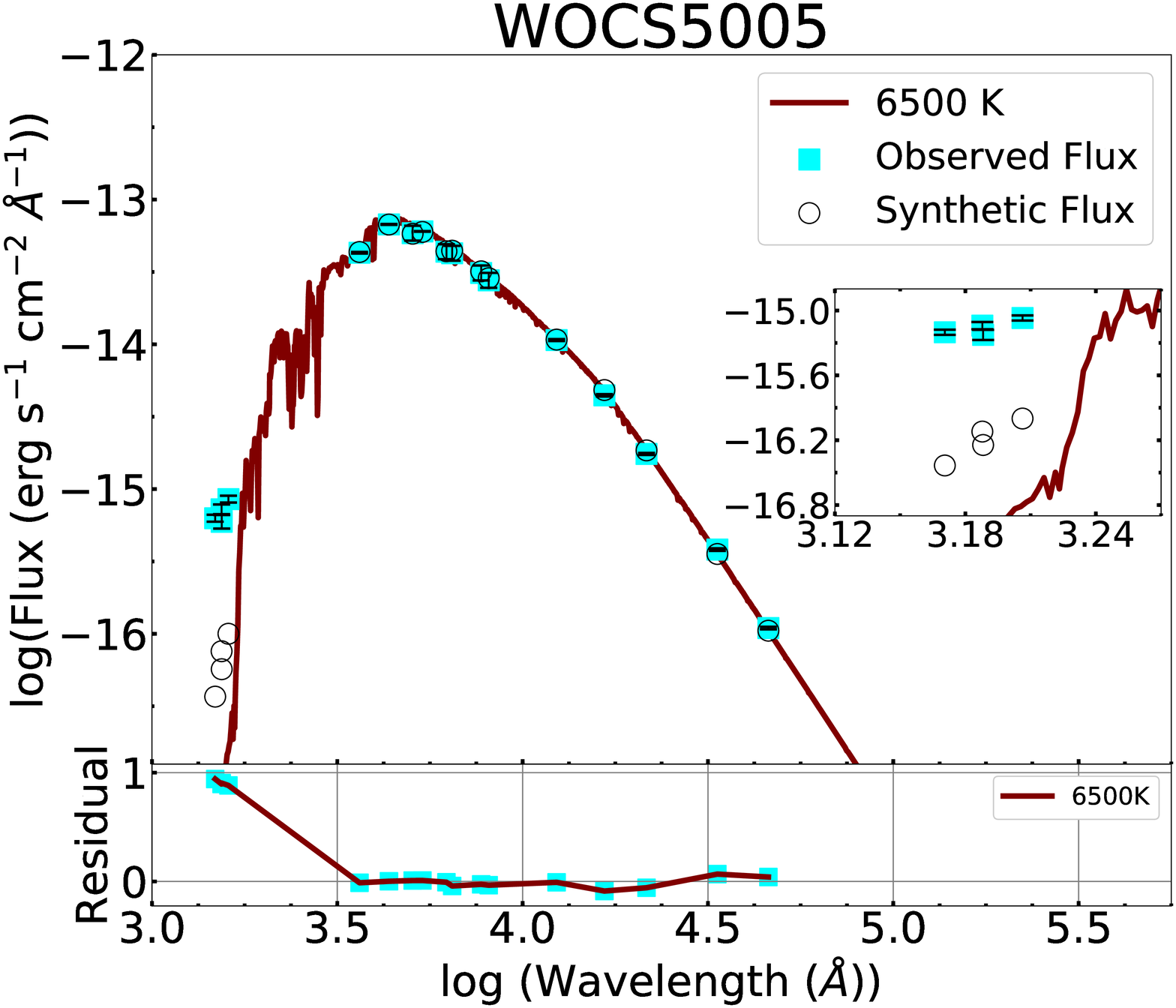}
\includegraphics[width=2.35in]{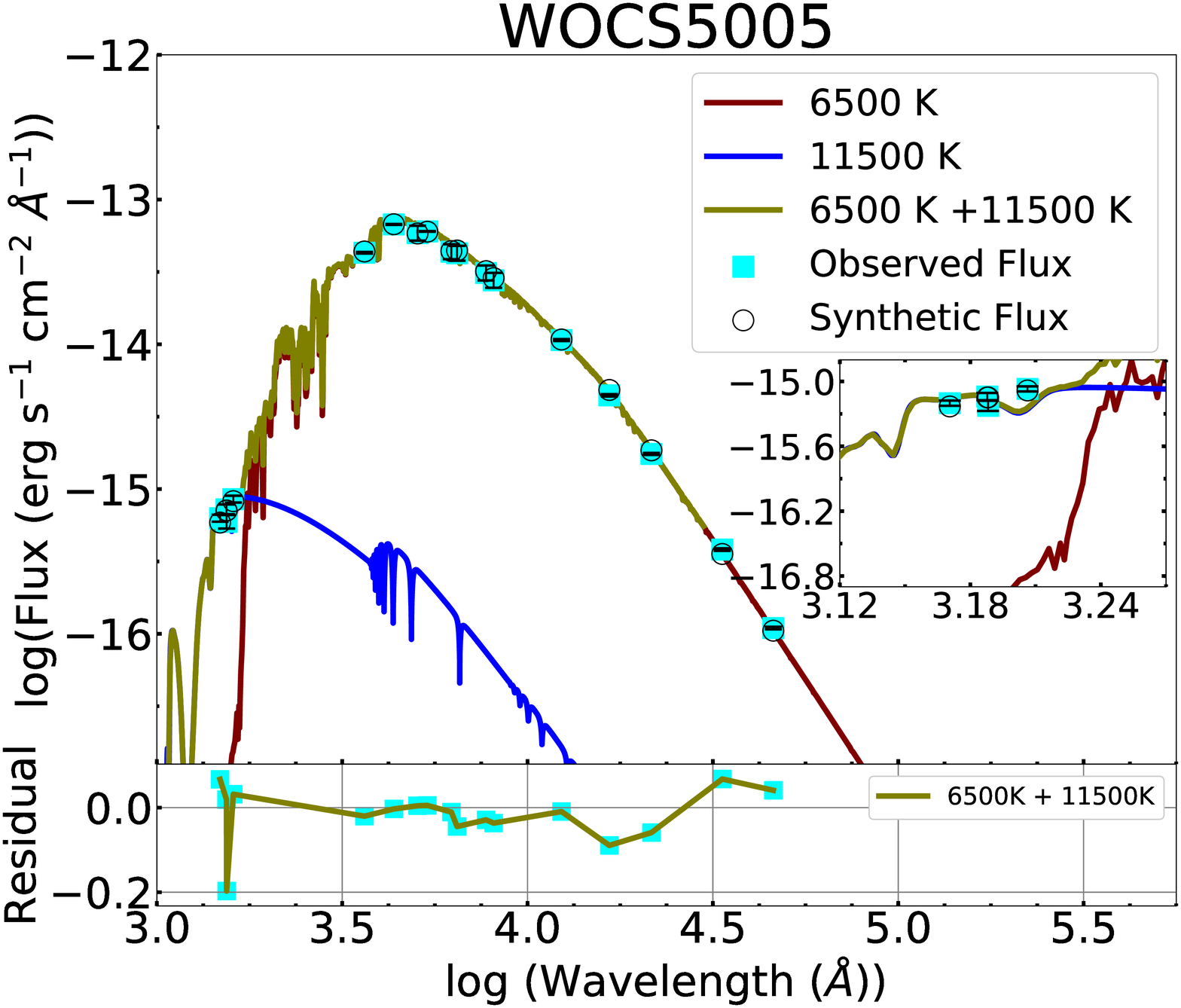}
\caption{Extinction Corrected SED fit of WOCS5005 : Single SED fit (Left) and Composite SED fit (Right). The cyan points indicate the observed flux and black open circle are the corresponding synthetic flux. Kurucz model spectra for the cool component (BSS) is shown in maroon, Koester WD model spectra for the hot component (WD) is shown in blue and the composite fit is shown in olive. The inset shows the zoomed UV region of the fit.}
 \label{fig1}
\end{center}
\end{figure}

The cluster M67, well known for the presence of several BSSs, has been put to detailed tests to understand BSS formation pathways. The main aim of this study is to check for the presence of hot companions to BSSs in this cluster. UVIT detected 10 BSSs of the 14 confirmed members of the cluster identified by \cite[Geller et~al. (2015)]{Geller2015} through radial velocity study. Here we restrict our sample to 5 BSSs that show significant excess UV flux in the residual single SED fit than expected for the BSSs corresponding to its T$_{eff}$, suggesting the presence of a hot component for these BSSs. The composite SEDs for these 5 BSSs are constructed by adding the flux of the hot component and the BSS at all wavelengths, in order to fit the observed full SED. We have used Kurucz model (\cite[Castelli et~al. 1997]{Castelli1997}) to fit the BSSs and Koester WD model (\cite[Koester 2010]{Koester2010}) to fit the hotter components, which are sub-luminous. In Fig. \ref{fig1}, we show the single and composite SED fits of one BSS (WOCS5005). The figure shows the scaled and best fitting Kurucz spectrum in maroon, Koester WD spectrum in blue and composite spectrum in olive along with the corresponding T$_{eff}$. The observed flux is shown in cyan square points and corresponding synthetic flux obtained by convolving with the respective filter are shown in black open circles. It is seen from this figure that the observed flux are in excess in the UV region in comparison to the expected flux for the BSS, thus requiring a double SED fit. 
Similar analysis were carried out for other 4 BSSs (WOCS1007, WOCS2013, WOCS3013 and WOCS4006) and they fit well with a composite SED fit. A detailed analysis of WOCS1007 is carried out by \cite[Sindhu et al. (2019)]{2019Sindhu} and they suggested that the hot companion of the BSS is a low mass He-WD of $\sim$ 0.2M$_{\odot}$.
\begin{figure}[htb]
\begin{center}
\includegraphics[width=2.35in]{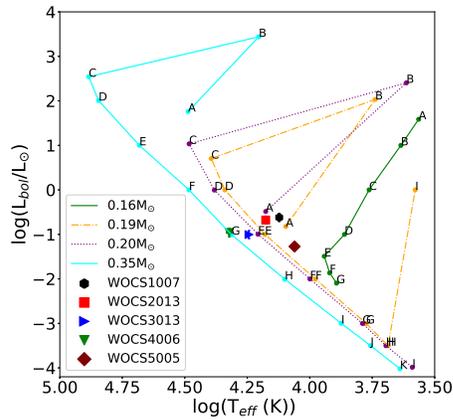}
\caption{H-R diagram of the hot companions of the 5 BSSs are shown along with the \cite[Panei et.~al. (2007)]{Panei2007} WD model taken from their Table 3. The label indicates the evolutionary phase of the WD of masses 0.19 to 0.35M$_{\odot}$, post binary evolution beginning from 'A'.}
 \label{fig2}
 \end{center}
\end{figure}

A H-R diagram of the hot companions of the BSSs is plotted in Fig. \ref{fig2} with their estimated temperature and luminosity obtained from the SED fits, along with \cite[Panei et.~al. (2007)]{Panei2007} WD model from their Table 3. The label in the Fig. \ref{fig2} indicates the evolutionary phases of the WD, post their binary evolution, which begins at `A'. Further details of these points can be found in their paper. The figure suggests that the WD companions of the 5 BSSs have a mass range of 0.2 to 0.35 M$_{\odot}$ and hence are probably low mass He-WD. As the formation of low mass He-WD would take more than a Hubble time, these are probably remnants of binary evolution.

\section{Conclusions}

We present the detection of possible WD companions to 5 bright BSSs in M67, imaged using FUV images from the UVIT. The multiwavelength SEDs covering 0.12-11.5$\mu$m range, were found to require binary spectral fits for 5 BSSs, consisting of a cool (BSS) and a hot companion. The parameters (Luminosity, T$_{eff}$ and Mass) of the hot companion suggest them to be low mass WDs, which are formed only in close binaries as a result of MT.

\section*{Acknowledgements}
UVIT project is a result of collaboration between IIA, IUCAA, TIFR, ISRO, and CSA. S.N acknowledges support from SERB for ITS grant to attend the IAUS 351 \& MODEST 19 and CSIR for grant 9/890(0005)/17 EMR-I.

\end{document}